\documentclass[12pt,preprintnumbers]{revtex4-2}
\usepackage{amsfonts}
\usepackage{amsmath}
\usepackage{amssymb}
\usepackage{graphicx}
\usepackage{multirow}
\usepackage[colorlinks=true]{hyperref}
\usepackage{caption}
\usepackage{subcaption}
\newcommand{\al}{\alpha}

\newcommand{\veps}{\varepsilon}

\newcommand{\rar}{\rightarrow}

\newcommand{\non}{\nonumber}

\newcommand{\re}[1]{(\ref{#1})}

\begin{document}

\title{The ground electronic state of CS: the potential curve and associated Born-Oppenheimer rovibrational spectrum}

\date{\today}

\author{Horacio~Olivares-Pil\'on}
\email{horop@xanum.uam.mx}
\affiliation{Departamento de F\'isica, Universidad Aut\'onoma Metropolitana-Iztapalapa,
Apartado Postal 55-534, 09340 M\'exico, D.F., Mexico}

\author{Alexander~V.~Turbiner}
\email{turbiner@nucleares.unam.mx}
\affiliation{Instituto de Ciencias Nucleares, Universidad Nacional
Aut\'onoma de M\'exico, Apartado Postal 70-543, 04510 M\'exico, D.F., Mexico}

\begin{abstract}
{
Basics of the Born-Oppenheimer (B-O) approximation are reviewed. Assuming the domain of applicability of B-O approximation is limited to 4 significant digits (s.d.) in energy spectrum, where mass, relativistic and QED corrections do {\it not} contribute, it is shown that for carbon monosulfide ${\rm C}\,{\rm S}$ the potential curve $V(R)$ for the electronic ground state $X^1\Sigma^+$ can be constructed analytically 
in the form of two-point Pade approximant $\frac{1}{R}\ P(5,10)(R)$ in the whole range of internuclear distances $R \in [0,\infty)$. Pade approximant is fixed by taking into account the turning points with 4 s.d. accuracy, found by Coxon and Hajigeorgiou \cite{CH:2023}, and asymptotics at small and large internuclear distances,

By solving two-body radial nuclear Schr\"odinger equation with the potential $V(R)$ (with standard centrifugal potential included) in the Lagrange Mesh method, the whole B-O rovibrational spectrum for ${}^{12} {\rm C}\,{}^{32} {\rm S}$ diatomic molecule (taken as a particular example) is found:
the $\sim 14562$ rovibrational energy states with angular momentum $L_{max}=289$ and vibrational quantum number $\nu_{max} \sim 82$ with accuracy $\sim 10^{-4}$\,{\it hartree} in energy. It is shown that the experimentally observed transition energies are reproduced within 3-5 s.d. Critical analysis of existing theoretical (phenomenological) results on the rovibrational spectrum is carried out and its comparison with present ones is made.}

\end{abstract}

\maketitle

%\maketitle

\section*{Introduction}
{
Needless to say that any atomic/molecular system is quantum many body Coulomb system with many degrees of freedom
made from nuclei and electrons. The celebrated Born-Oppenheimer (B-O) approximation (in other words, the
static approximation) provides invaluable tool to consider such a system.
Technically, the B-O approximation in the molecular physics is based on splitting
the non-relativistic quantum many-body Coulomb problem into the electronic and nuclear parts.
In the electronic part the nuclei are considered as infinitely-massive, the internuclear distances $\{R\}$
are fixed, since they are non-dynamical (classical) variables, the spectrum of the electronic Hamiltonian,
which is sum of kinetic energy of electrons and the potential of the Coulomb interactions, is calculated.
The eigenvalues/the electronic energies are found as a function of the internuclear distances $\{R\}$,
they are called the potential curves/surfaces. Note that any classical internuclear distance $R$ is
not integral of motion of the total Hamiltonian, it is not constant of motion, it is a conserved quantity 
of the electronic Hamiltonian only.
In total Hamiltonian the $R$ becomes a dynamical variable, it should be well-described by the Gaussian
distribution with width $\sigma \sim \sqrt{1/M_{nuclei}}$. Internuclear distances $\{R\}$ play a role of
external parameters/arguments of the spectra of electronic energies. Thus, the electronic Hamiltonian
defines multi-center quantum problem, it describes the so-called Quantum Mechanics of Coulomb
Charges (QMCC). Finding the eigenvalues of the electronic Hamiltonian is usually a subject
of the so-called {\it ab initio} calculations. For 1-2 electron cases it is relatively simple technical
problem which becomes hardly-treatable for many-electron case. In a very few cases the spectra
of electronic energies can be found parameterically, as a continuous function of internuclear distances.
In general, there is quite non-trivial technical issue:
{\it ab initio} studies provide the eigenvalues at discrete values of the internuclear distances,
but how to make interpolation between these points and extrapolation outside, towards large and
small internuclear distances? It is not always clear. It hints that perturbation theory results
at small internuclear distances and multipole expansion at large distances should be added to
{\it ab initio} results.
In the case of atoms we have one-center problem as the realization of
the static approximation in the atomic physics. In the nuclear part the nuclear Hamiltonian
is formed as sum of kinetic energy of nuclei and electronic energy (potential curve/surface)
as a potential. In the case of two heavy centers (diatomic molecule)
the spectrum of the nuclear Hamiltonian is called the ro-vibrational spectrum.
The nuclear Hamiltonian is given by the radial Hamiltonian with added centrifugal term.

The important question related to the B-O approximation is its accuracy, which defines its domain of applicability.
Naturally, there are three main types of corrections: (i) nuclear mass corrections
$\sim m_e/M_{nuclei}$, (ii) relativistic corrections $\sim (v/c)^2$ and (iii)
QED corrections $\sim \al^2$.
In studies of the helium-like and lithium-like ions with nuclear charge $Z \leq 20$
it was shown that the finite nuclear mass effects do not change 4–5 s.d., 
and the leading relativistic and QED effects leave
unchanged 4 s.d. in the ground state energy of these ions\cite{TLO:2019}. Hence, the first 4-5 s.d.
in the energy are correction-free. There are indications that the similar observation
holds for few-electron diatomic molecules, see e.g. \cite{Komasa:2019} and references 
therein. { We assume the existence of correction-free domain for any atomic system, for 
arbitrary diatomic molecule.} All that defines the domain of applicability
of the B-O/static approximation. For convenience we assume conventionally
that the first {(four figures) $\equiv$ (four significant digits) in the total energies are 
correction-free, they can be reproduced in the B-O formalism. 
It implies that the potential curve/surface is defined consistently within 4 s.d. {\it only}.

For the case of diatomic molecules there exists an alternative approach by solving
the two-body nuclear Schr\"odinger equation employing the so-called {\it inverse problem method}.
Taking the available experimental data for the energy spectrum one can restore the corresponding
turning points on the potential curve. It relies on the so-called Rydberg-Klein-Rees (RKR)
method and its modifications. This method is based on the Bohr-Sommerfeld quantization rule and
provides ``astonishingly accurate results" (A. Mantz, et al. J. Mol. Spectrosc. 39, 180 (1971))
%\footnote
{(Its justification is not completely clear to the present authors.)}.
Then the potential curve is modelled by a parametric potential whose parameters are found
by fitting the turning points, where the choice of the modelling potential plays a crucial role
for getting adequate rovibrational spectra. It must be emphasized that insufficiency of available experimental data can also lead to a loss of accuracy and even missing parts of the rovibrational spectrum.
The best example of this situation appears in a study of hydrogenic halides and their isotopologues, which was carried out by Coxon and Hajigeorgiou in \cite{CH:2015}: 
for (H,D,T)Cl, see \cite{OT:2023} and for (H,D,T)Br, (H,D,T)I, see \cite{AdT:2024} - thousands of 
B-O rovibrational states are missing, while all vibrational states are present together with some rovibrational states! This alternative approach allows to avoid solving the electronic Schr\"odinger equation. However, in order to reach spectroscopical accuracies of 6-7 s.d. in (transition) energies (in absence of reliable data on mass, relativistic and QED corrections) it is proposed to modify phenomenologically the nuclear Schr\"odinger equation by introducing factors in front of kinetic energy and centrifugal potential as well as the reduced mass dependence into the potential curve = the so-called electronic term, see e.g. \cite{CH:2015}. It allows to reproduce the available experimental data on transition frequencies/energies but it {\it does NOT guarantee} the same quantitative description for so-far-unavailable experimental data. It is typical situation for phenomenological approach.\\ 
}
\indent
Carbon monosulfide CS is a short-lived diatomic molecule (sometimes, it is called dimer) ~\cite{WR:1973}
of great astrophysical importance. CS was the first sulfur-containing molecule detected
in interstellar media~\cite{PSWJ:1971} and has been identified in a wide variety
of astrophysical objects (see e.g. \cite{MKS:1988} and recent articles
\cite{Paulose:2015, PCSFBMM:2018} and the references therein).
In this article, in the true B-O approximation, an analytical representation
for the potential energy curve for the CS molecule is proposed in the form of two-point
Pad\'e approximant. It is based on the methodology formulated in~\cite{TO:2022}:
the perturbation theory in small $R \rightarrow 0$ and the multipole expansion at
large $R\rightarrow\infty$ are join into a single function specified by taking into account 
the Rydberg-Klein-Rees(RKR)-style turning points derived from experimental data in 
\cite{CH:2023}.
Then this potential curve, modified by centrifugal potential, is used as the potential
in the nuclear two-body radial Schr\"odinger equation. In particular, it allows to check 
the consistence of the experimental energy spectrum used to derive the RKR-style turning 
points with the spectrum found of the nuclear radial Schr\'odinger equation.
This procedure has been successfully applied to several neutral heteronuclear molecules
LiH~\cite{TO:2022}, HeH~\cite{OT:2018}, ClF~\cite{OT:2022}, H(D,T)F~\cite{AG-OP:2022},
as well as (H,D,T)Cl, see \cite{OT:2023} and (H,D,T)Br, (H,D,T)I, see \cite{AdT:2024}, 
making it possible to calculate the whole rovibrational energy spectrum with correct 3-5
figures, when compared to experimental or theoretical results. This accuracy
in energy is consistent with definition of the correction-free domain of B-O approximation: 
mass corrections, relativistic corrections, QED corrections do not influence the first 
3-5 figures in the rovibrational energies.\\
\indent
Throughout the paper the distances are given in atomic units (a.u.), while the energies are
in {\it hartree}.

\section{Potential energy curve}

The diatomic molecule of carbon monosulfide CS is made of two neutral atoms:
the Carbon ($Z_{\rm C}=6$) and the Sulfur ($Z_{\rm S}=16$), it contains 22 electrons. When the nuclear masses 
are assumed infinite, the potential energy curve appears (in the B-O 
approximation) as the eigenvalue of the electronic Hamiltonian.
The B-O potential energy curve $E_d(R)$ for the ground electronic state
of the molecular system CS is usually related to the total energy $E_{total}$ as
\begin{equation}
    E_d(R)\ =\ E_{total}(R)\ -\ (E_{\rm C}+E_{\rm S})\ ,
\end{equation}
where the ground state energies of the Carbon (C) and Sulfur (S) atoms are
$E_{\rm C}=-37.855787$~{\it hartree} and  $E_{\rm S}=-399.08$~{\it hartree}~\cite{NIST:2022},
respectively. In the united atom limit $R \rar 0$, the single nucleus of the CS molecule
corresponds to the Titanium (Ti), where their ground state energy is
$E_{\rm Ti}=-853.36$~{\it hartree}~\cite{NIST:2022}.

At small internuclear distances $R \rar 0$, the potential energy curve is given by the 
pertubation theory expansion~\cite{Bingel:1958}
\begin{equation}
\label{EsmallR}
  E_d\ =\ \frac{Z_{\rm C}Z_{\rm S}}{R}\ +\ \veps_0\ +\ 0 \cdot R\ +\ O(R^2)\ ,
\end{equation}
where the first term $Z_{\rm C}Z_{\rm S}/R$ represents the Coulomb repulsion
and $\veps_0=E_{\rm Ti} +|E_{\rm C}+E_{\rm S} |=-416.424213$~{\it hartree}.
On the other hand the behavior of the potential energy at large internuclear
distances $R \rightarrow \infty$ corresponds to the multipole expansion of the
(induced) moments
\begin{equation}
\label{ElargeR}
  E_d\ =\  -\frac{C_5}{R^5} - \frac{C_6}{R^6}\ -\ \frac{C_8}{R^8}\ +\ \cdots \ .
\end{equation}
For neutral heteronuclear diatomic molecules, $C_5=0$~\cite{MK:1971,IK:2006} and
the van der Waals constant is $C_6=41.95229$~\cite{CH:2023} 
%\footnote{
(Estimate from the London formula gives $C_6=58.02$~\cite{PCSFBMM:2018}).
Due to the fact that the ground states of the atoms, to which CS dissociates,
appear to be in a $^3$P state~\cite{CH:2023,MSE:2018}, the constant $C_5$ can be
non-zero and it can be taken as $C_5=28.70673$~\cite{CH:2023} 
\footnote{Alternative value is $C_5=27.34$ proposed in~\cite{PCSFBMM:2018}}.
We will explore both possible scenarios, $C_5=0$ and $C_5\neq 0$.

Matching the expansions (\ref{EsmallR}) and (\ref{ElargeR}) into a single function
leads to the B-O potential energy curve in the form of a two-point Pad\'e approximant,
\begin{equation}
\label{Ed}
     E_d(R)\ =\ \frac{1}{R}\,{\rm Pade}[N /M](R)_{n_0,n_{\infty}}\ \equiv \
     \frac{1}{R}\, \frac{P_N}{Q_{M}}\ ,
\end{equation}
where $P_N, Q_M$ are polynomials in $R$ of degree $N$ and $M$, respectively,
and $n_0$/$n_{\infty}$ are the numbers of coefficients in the expansion
(\ref{EsmallR})/(\ref{ElargeR}) which we want to reproduce exactly.
It is important to note that for positive $R > 0$ we require $Q(R)>0$, thus,
denominator in (\ref{Ed}) can have either negative, $R < 0$, and/or complex roots {\it only}.
As a result of taking two different asymptotic expansions at large internuclear distances
in (\ref{ElargeR}), two different analytical expressions emerge
as a result of matching:
%%----------------------------------------------------------------------------------------
%\textcolor{blue}
{
\begin{itemize}
\item[($i$)] Assuming $C_5=0$ with $N=5$ and $M=10$, the potential energy curve is represented as
\begin{equation}
\label{fitP49}
    E_{\{3,2\}}^{C_6}(R)\ =\
    \frac{1}{R}\ \frac{Z_{\rm C}Z_{\rm S}\,+\,a_1 R\,+a_2 R^2\,+ a_3 R^3\,+a_4 R^4\,-\,a_5 R^5}
    {1+\al_1 R + \al_2 R^2\, +\,\sum_{i=3}^{8}b_i R^i -\al_3 R^9+ b_{10}\, R^{10}}\ \ ,
\end{equation}
where four parameters
\begin{eqnarray}
\label{par-1}
     \al_1 & =  & (a_1-\veps_0)/Z_{\rm C}Z_{\rm S}\ ,\non \\
     \al_2 & =  & (\veps_0^2 + Z_{\rm C}Z_{\rm S}\, a_2  - a_1\veps_0)/(Z_{\rm C}Z_{\rm S})^2\ ,\non \\
     \al_3 & =  & a_4/C_{6}\ ,\non \\
     a_5 & =  & b_{10}\, C_6\ ,
\end{eqnarray}
are constrained in order to guarantee that the three coefficients in front of the $R^{-1}$, $R^0$ and $R$
terms at small internuclear distance expansion~\re{EsmallR} and the two coefficients in
front of $R^{-6}$ and $R^{-7}$ at large one~\re{ElargeR} are reproduced exactly. 
The remaining 12 free parameters in (\ref{fitP49}) are fixed by making fit of data for turning points 
(on a potential curve) presented in~\cite{CH:2023} in domain $R \in (2.3,4.5)$\,a.u. 
and the 9 turning points in the domain $R \in (5.0,7.5)$\,a.u., extracted from the potential 
energy curve presented in~\cite{CH:2023}. The explicit values of the parameters are the following
\begin{equation}
\begin{array}{lrrrrr}
a_1 = &  14.30216\,, & a_5 = &  {0.197470091}\,, & b_6 = &   -3.850841\,, \\
a_2 = &  12.96279\,, & b_3 = & -14.81227\,,  & b_7    = &   0.1916986\,, \\
a_3 = & -22.19022\,, & b_4 = & -3.677174\,,  & b_8    = &   0.2344471\,, \\
a_4 = &  2.479577\,, & b_5 = &  10.19328\,,  & b_{10} = & \ \ \ 0.0047070159 \ .\\
\end{array}
\label{par-2}
\end{equation}
The potential curve \re{fitP49} with parameters (\ref{par-1}) - (\ref{par-2})
leads eventually to the 3-4 s.d. in experimentally-observed vibrational energies (see below).
The minimum of the potential energy curve given by~\re{fitP49} is characterized by
$R_{min}=2.90066$\,a.u. and $E_{min}=-0.27317$\,hartree in close
agreement with results from ~\cite{CH:2023}:
$R_{equilibrium}=2.90062$\,a.u. and $E_{equilibrium}=-0.27319$\,hartree.
\item[($ii$)] Taking $C_5=28.70673$, presented in~\cite{CH:2023}, see (\ref{ElargeR}), 
and choosing $N=5$, $M=9$ the potential curve appears in the form
\begin{equation}
\label{fitP59}
    E_{\{3,3\}}^{C_5}(R)\ =\
    \frac{1}{R}\ \frac{Z_{\rm C}Z_{\rm S}\,+\,a_1 R\,+a_2 R^2\,+ a_3 R^3\,+a_4 R^4\,-\,a_5 R^5}
    {1+\al_1 R + \al_2 R^2\, +\,\sum_{i=3}^{6}b_i R^i+\al_3 R^7 -\al_4 R^8+ b_9\, R^9} \,.
\end{equation}
By imposing constraints
\begin{eqnarray}
\label{parc5}
     \al_1 & = & (a_1-\veps_0)/Z_{\rm C}Z_{\rm S}\ ,
\non \\
     \al_2 & = & (\veps_0^2 + Z_{\rm C}Z_{\rm S}\, a_2  - a_1\veps_0)/(Z_{\rm C}Z_{\rm S})^2\ ,
\non \\
     \al_3 & = & (b_9\,C_6 (a_4+b_9\,C_6)-a_3\,a_5)/(a_5\,C_5)\ ,
\non \\
     \al_4 & = & (b_9\,C_6+a_4)/C_5\ ,
\non \\
      a_5  & = &  b_9\, C_6\ ,
\end{eqnarray}
make the first three coefficients in~\re{EsmallR} ($R^{-1}$, $R^0$, $R$) and in~\re{ElargeR} 
($R^{-5}$, $R^{-6}$, $R^{-7}$) be correctly reproduced.
The 10 remaining free parameters in (\ref{fitP59}) are fixed by making fit by using
the same data presented in~\cite{CH:2023} on turning points for domain $R \in (2.3,4.5)$\,a.u. 
and the 9 turning points in the domain $R \in (5.0,7.5)$\,a.u., 
as was used in $E_{\{3,2\}}^{C_6}(R)$ see~\re{fitP49}. Their explicit values of parameters are
\begin{equation}
\begin{array}{lrrrrr}
a_1 =&   12.15042\,, & a_5 = &  0.02828226848\,, & b_6 = & 2.207317\,, \\
a_2 =&  -38.07173\,, & b_3 = & -33.54668\,,   & b_9 = & 0.000985213916\,, \\
a_3 =&   6.220700\,, & b_4 = &  26.36060\,,   & &  \\
a_4 =& -0.1047569\,, & b_5 = & -10.311475\,.  & &  \\
\end{array}
\label{par-3}
\end{equation}
cf. (\ref{par-2}).
%\begin{equation}
%\begin{array}{lrrrrr}
%a_1 =& -17.85647\,, & a_5 = &   0.1350201\,, & b_6 = & 0.2073629\,, \\
%a_2 =& -6.366344\,, & b_3 = & -30.18571\,,   & b_9 = & 0.00470343\,, \\
%a_3 =& -5.332298\,, & b_4 = &  20.73868\,,   & &  \\
%a_4 =&  1.647479\,, & b_5 = &  -5.768915\,.  & &  \\
%\end{array}
%\label{par-3}
%\end{equation}
This fit leads to the turning point positions, found in \cite{CH:2023} in 3-4 s.d.
The minimum of the potential energy curve is calculated from~\re{fitP59}:
the obtained values $R_{min}=2.90060$\,a.u. and 
$E_{min}=-0.27317$~{\it hartree} coincide with $R_{equilibrium}=2.90062$\,a.u. 
and $E_{equilibrium}=-0.27319$~{\it hartree} found in \cite{CH:2023}.
\end{itemize}
}
%%----------------------------------------------------------------------------------------

Table ~\ref{TcompV} displays the potential energy curve $E(R)$ obtained from
the analytic expressions~\re{fitP49} and ~\re{fitP59}, denoted as $E^{C_6}$ and $E^{C_5}$,
respectively. These curves are compared with data from~\cite{CH:2023}.
As can be seen, except for a few points, that there is a coincidence in 4 decimal digits (d.d.).
Fig.~\ref{FVpotCS}  depicts the analytical curve of potential energy as a function
of internuclear distance $R$ together with turning points
from~\cite{CH:2023}. Note that there is no visible difference between the analytical
curves~\re{fitP49} and~\re{fitP59} in the domain $R \lesssim 9$\, a.u., the differences occur for 
$R > 9$\, a.u. which defines highly-excited, weakly-bound states.

\begin{table}[!thb]
\caption{Potential energy curve $E$ for the electronic ground state
$X^1\Sigma^+$ for the CS diatomics as a function of the internuclear distance $R$
in $R<R_{eq}=2.9006$~a.u. and $R>R_{eq}$.
The second and sixth columns are from~\cite{CH:2023}. For a given internuclear distance
$R$:  $E^{C_6}$ and $E^{C_5}$ are from~\re{fitP49} and ~\re{fitP59}, respectively. In general,
they coincide in 4 d.d.: $E^{C_5}$ data are shown only when they differ from $E^{C_6}$. 
{Data by bold extracted from the analytical potential curve presented in~\cite{CH:2023}.}}
\begin{center}
\scalebox{0.9}{%
\begin{tabular}{c| ccc| c| ccc}
\hline\hline
$R_{min}$& $E^{R_{min}}$~\cite{CH:2023} & $E^{C_6}$~\re{fitP49}&$E^{C_5}$~\re{fitP59}&
$R_{max}$& $E^{R_{max}}$~\cite{CH:2023} & $E^{C_6}$~\re{fitP49}& $E^{C_5}$~\re{fitP59}\\
\hline
2.9006& -0.2732& -0.2732&        &       &        &        &        \\
2.8021& -0.2703& -0.2703& & 3.0098& -0.2703& -0.2703& -0.2702\\
%2.7361& -0.2645& -0.2645& 	 & 3.0974& -0.2645& -0.2645& 	    \\
2.6934& -0.2587& -0.2587& 	 & 3.1618& -0.2587& -0.2587& 	    \\
2.6603& -0.2531& -0.2531& 	 & 3.2169& -0.2531& -0.2531& 	    \\
%2.6328& -0.2474& -0.2474& 	 & 3.2666& -0.2474& -0.2474& 	    \\
%2.6090& -0.2419& -0.2419& 	 & 3.3128& -0.2419& -0.2419& 	    \\
2.5879& -0.2364& -0.2364& 	 & 3.3564& -0.2364& -0.2364& 	    \\
%2.5690& -0.2309& -0.2309& 	 & 3.3982& -0.2309& -0.2309& 	    \\
%2.5517& -0.2255& -0.2256& 	 & 3.4384& -0.2255& -0.2256& 	    \\
2.5359& -0.2202& -0.2202& 	 & 3.4775& -0.2202& -0.2202& 	    \\
%2.5212& -0.2150& -0.2150& 	 & 3.5157& -0.2150& -0.2150& 	    \\
%2.5075& -0.2097& -0.2097&-0.2098& 3.5531& -0.2097& -0.2097& -0.2098\\
2.4948& -0.2046& -0.2046& 	 & 3.5899& -0.2046& -0.2046& 	    \\
%2.4827& -0.1995& -0.1995& 	 & 3.6261& -0.1995& -0.1995& 	    \\
%2.4714& -0.1945& -0.1945& 	 & 3.6620& -0.1945& -0.1945& 	    \\
2.4606& -0.1895& -0.1895& 	 & 3.6974& -0.1895& -0.1895& 	    \\
2.4504& -0.1846& -0.1846& 	 & 3.7326& -0.1846& -0.1846& 	    \\
%2.4407& -0.1797& -0.1797& 	 & 3.7676& -0.1797& -0.1797& 	    \\
%2.4315& -0.1749& -0.1749& 	 & 3.8024& -0.1749& -0.1749& 	    \\
%2.4226& -0.1701& -0.1701& 	 & 3.8371& -0.1701& -0.1701& 	    \\
%2.4142& -0.1655& -0.1655&-0.1654& 3.8717& -0.1655& -0.1654& 	    \\
%2.4061& -0.1608& -0.1608& 	 & 3.9063& -0.1608& -0.1608& 	    \\
2.3983& -0.1563& -0.1562& 	 & 3.9408& -0.1563& -0.1562& 	    \\
%2.3908& -0.1517& -0.1517& 	 & 3.9753& -0.1517& -0.1517& 	    \\
2.3836& -0.1473& -0.1473& 	 & 4.0099& -0.1473& -0.1473& 	    \\
%2.3767& -0.1429& -0.1429& 	 & 4.0446& -0.1429& -0.1429& 	    \\
2.3700& -0.1385& -0.1385& 	 & 4.0793& -0.1385& -0.1385& 	    \\
%2.3636& -0.1342& -0.1342& 	 & 4.1142& -0.1342& -0.1342& 	    \\
2.3574& -0.1300& -0.1300& 	 & 4.1492& -0.1300& -0.1300& 	    \\
%2.3514& -0.1258& -0.1258& 	 & 4.1844& -0.1258& -0.1258& 	    \\
%2.3456& -0.1217& -0.1217& 	 & 4.2198& -0.1217& -0.1217& 	    \\
2.3400& -0.1177& -0.1177& 	 & 4.2554& -0.1177& -0.1177& 	    \\
2.3346& -0.1137& -0.1137& 	 & 4.2913& -0.1137& -0.1137& 	    \\
2.3294& -0.1097& -0.1097& 	 & 4.3275& -0.1097& -0.1097& 	    \\
2.3243& -0.1058& -0.1058& 	 & 4.3639& -0.1058& -0.1058& 	    \\
%2.3194& -0.1020& -0.1020& 	 & 4.4006& -0.1020& -0.1020& 	    \\
2.3146& -0.0982& -0.0982& & 4.4378& -0.0982& -0.0983& 	    \\
%2.3100& -0.0945& -0.0945& 	 & 4.4752& -0.0945& -0.0945& 	    \\
%2.3055& -0.0909& -0.0909& 	 & 4.5131& -0.0909& -0.0909& 	    \\
2.3012& -0.0873& -0.0873& 	 & 4.5515& -0.0873& -0.0873& 	    \\
      &       &    &   & {\bf 5.0000}&{\bf-0.0526}& -0.0526&	    \\
%      &       &    &   & {\bf 5.2000}&{\bf-0.0412}& -0.0411& -0.0412\\
      &       &    &   & {\bf 5.4000}&{\bf-0.0319}& -0.0319&	    \\
%      &       &    &   & {\bf 5.6000}&{\bf-0.0245}& -0.0245&	    \\
      &       &    &   & {\bf 5.8000}&{\bf-0.0187}& -0.0187&	    \\
      &       &    &   & {\bf 6.0000}&{\bf-0.0143}& -0.0143& -0.0142\\
      &       &    &   & {\bf 6.5000}&{\bf-0.0072}& -0.0072&	    \\
      &       &    &   & {\bf 7.0000}&{\bf-0.0037}& -0.0037& -0.0038\\
      &       &    &   & {\bf 7.5000}&{\bf-0.0020}& -0.0020& -0.0021\\
\hline\hline
\end{tabular}}
\end{center}
\label{TcompV}
\end{table}

\begin{figure}[h!]
\includegraphics[scale=2.0]{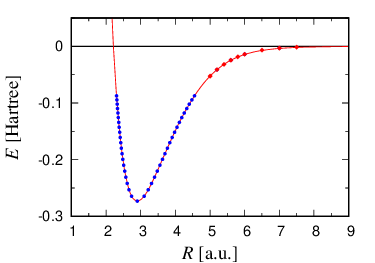}
\caption{Potential energy $E$ for the ground state $X^1\Sigma^+$ for the
   carbon monosulfide CS {\it vs.} internuclear distance $R$:
   $(i)$ the potential curves produced by $E^{C_6}$~\re{fitP49} and by
   $E^{C_5}$~\re{fitP59} {\it coincide} (red solid line) in domain $R < 10$\,a.u., 
   $(ii)$ data~from~\cite{CH:2023} (blue bullets) $(iii)$
   {9 points in $R \in (5.0,7.5)$\,a.u. extracted from the potential 
    energy curve proposed in~\cite{CH:2023} (red diamonds).} }
\label{FVpotCS}
\end{figure}

Concluding we must emphasize that the constructed B-O potential curves for CS do
{\it not} depend on the isotopologue. This feature was already seen in the results
presented in \cite{CH:2023}: for all eight isotopologues studied there the equilibrium 
distances coincides within five significant figures, see Table V therein.

%--------------------------------------------------------------------------------------------------------------
\section{B-O Rovibrational Spectra}

In the B-O approximation, the rovibrational spectrum
of the carbon monosulfide CS  is found by solving the nuclear radial Schr\"odinger equation
\begin{equation}
\label{NucSc}
\left[-\frac{1}{2\mu}\frac{d^2}{dR^2}\ +\ \frac{L(L+1)}{2\mu R^2}\ +\ V(R)\right]\,\phi(R)\
=\  E_{(\nu,L)}\, \phi(R)\ ,
\end{equation}
where  $\mu=m_{\rm C} m_{\rm S}/(m_{\rm C}+ m_{\rm S})$ is the reduced mass for the carbon
and sulfur nuclei, $L$ is the total angular momentum and $E_{(\nu,L)}$ is the rovibrational
energy of the state with vibrational and rotational quantum
numbers $\nu$ and $L$, respectively. The nuclear potential $V(R)=E_d$ is the electronic
energy curve given by~\re{fitP49} or~\re{fitP59}. The nuclear masses are chosen as 
$m_{\rm C} =21868.66183$~\cite{AWT:2003} for $^{12}$C and $m_{\rm S}=58265.52010$ for 
$^{32}$S~\cite{AWT:2003}, measured in the mass of the electron. Equation~\re{NucSc} is solved
by using the Lagrange-Mesh Method~\cite{DB:2015}.

The vibrational spectrum $E_{(\nu,0)}$  was obtained by considering the two expressions
for the potential curve, $V(R)=E^{C_6}$~\re{fitP49} and $V(R)=E^{C_5}$~\re{fitP59}.
For $\nu \in [0,39]$ the results for both cases are presented in Table~\ref{TvSClF}
together with experimental data reported in~\cite{CH:2023}.
When compared these first 40 vibrational energies $E_{(\nu,0)}$ from~\cite{CH:2023},
the absolute deviation $\sim 10^{-4}$\,~{\it hartree} from experimental data occurs 
for both theoretical curves, whenever $V(R)=E^{C_6}$~\re{fitP49} or 
$V(R)=E^{C_5}$~\re{fitP59} are taken. 
For large vibrational quantum numbers $\nu > 39$ the experimental data are absent, 
these energies are not taken into consideration in \cite{CH:2023}, they are not 
predicted there as well as the maximal vibrational quantum number $\nu_{max}$ is not 
indicated.

It is evident that the presence of the $C_5/R^5$ term in \re{fitP59} or its absence 
in \re{fitP49} should modify the energies of the vibrational states close 
to the dissociation limit only -  
in concrete calculations it was found that occurs when $\nu > 72$. It was also found that
for the ground electronic state X$^1\Sigma^+$ of the concrete diatomic molecule 
$^{12}$C$^{32}$S, it supports 83/85 vibrational states whether approximation $E^{C_6}$\re{fitP49}/$E^{C_5}$\re{fitP59} is assumed, 
hence, $\nu_{max}=82/84$, respectively. 
It is in a close agreement with $\nu_{max}=85 \rar \nu_{max}=84$,
obtained in \cite{PCSFBMM:2018}. 

It has to be emphasized that for $\nu > 80$ for all eight isotopologues, 
mentioned in \cite{CH:2023}, we found the vibrational energies appeared to be 
of $\lesssim 10^{-4}$~{\it hartree}: 
those weakly-bound states should be excluded from the consideration since mass effects 
could be crucial for their existence. In general, the comparison of the vibrational 
energy spectrum $E_{(\nu,0)}$ obtained with $E^{C_6}$\re{fitP49} and with $E^{C_5}$\re{fitP59}
the absolute deviation $\lesssim 10^{-4}$~{\it hartree} in the energy spectra occurs.

\begin{table}[!thb]
\caption{Vibrational energies $E_{(\nu,0)}$ for the electronic ground state 
$X^1\Sigma^+$ of the carbon monosulfide $^{12}$C$^{32}$S. Vibrational energies
from~\cite{CH:2023} presented in the 2nd and 6th columns. $E_{(\nu,0)}^{C_6}$ and
$E_{(\nu,0)}^{C_5}$ are the vibrational energies based on the potentials~\re{fitP49} and
~\re{fitP59}, respectively.}
\begin{center}
\scalebox{0.9}{
\begin{tabular}{r|  ccc |r ccc|r cc| r cc}
\hline\hline
$\nu$ &\cite{CH:2023} & $E_{(\nu,0)}^{C_6}$ & $E_{(\nu,0)}^{C_5}$ & $\nu$&\cite{CH:2023} &
$E_{(\nu,0)}^{C_6}$ & $E_{(\nu,0)}^{C_5}$ & $\nu$ & $E_{(\nu,0)}^{C_6}$ & $E_{(\nu,0)}^{C_5}$ &
$\nu$ & $E_{(\nu,0)}^{C_6}$ & $E_{(\nu,0)}^{C_5}$\\
\hline
0 & -0.2703& -0.2703& -0.2702& 20& -0.1655& -0.1654& -0.1654& 40& -0.0837& -0.0837& 60& -0.0263& -0.0263\\
1 & -0.2645& -0.2645& -0.2645& 21& -0.1608& -0.1608& -0.1608& 41& -0.0803& -0.0803& 61& -0.0241& -0.0241\\
2 & -0.2587& -0.2587& -0.2587& 22& -0.1563& -0.1562& -0.1562& 42& -0.0769& -0.0769& 62& -0.0220& -0.0220\\
3 & -0.2531& -0.2530& -0.2530& 23& -0.1517& -0.1517& -0.1517& 43& -0.0735& -0.0735& 63& -0.0200& -0.0200\\
4 & -0.2474& -0.2474& -0.2474& 24& -0.1473& -0.1473& -0.1473& 44& -0.0702& -0.0702& 64& -0.0181& -0.0181\\
5 & -0.2419& -0.2419& -0.2419& 25& -0.1429& -0.1429& -0.1429& 45& -0.0670& -0.0670& 65& -0.0163& -0.0163\\
6 & -0.2364& -0.2364& -0.2364& 26& -0.1385& -0.1385& -0.1385& 46& -0.0638& -0.0638& 66& -0.0146& -0.0145\\
7 & -0.2309& -0.2309& -0.2309& 27& -0.1342& -0.1342& -0.1342& 47& -0.0607& -0.0607& 67& -0.0129& -0.0129\\
8 & -0.2255& -0.2255& -0.2255& 28& -0.1300& -0.1300& -0.1300& 48& -0.0576& -0.0577& 68& -0.0113& -0.0113\\
9 & -0.2202& -0.2202& -0.2202& 29& -0.1258& -0.1258& -0.1258& 49& -0.0547& -0.0547& 69& -0.0099& -0.0099\\
10& -0.2150& -0.2149& -0.2149& 30& -0.1217& -0.1217& -0.1217& 50& -0.0517& -0.0518& 70& -0.0085& -0.0085\\
11& -0.2097& -0.2097& -0.2097& 31& -0.1177& -0.1176& -0.1176& 51& -0.0489& -0.0489& 71& -0.0072& -0.0072\\
12& -0.2046& -0.2046& -0.2046& 32& -0.1137& -0.1136& -0.1136& 52& -0.0461& -0.0461& 72& -0.0060& -0.0060\\
13& -0.1995& -0.1995& -0.1995& 33& -0.1097& -0.1097& -0.1097& 53& -0.0434& -0.0434& 73& -0.0050& -0.0050\\
14& -0.1945& -0.1944& -0.1945& 34& -0.1058& -0.1058& -0.1058& 54& -0.0407& -0.0407& 74& -0.0040& -0.0040\\
15& -0.1895& -0.1895& -0.1895& 35& -0.1020& -0.1020& -0.1020& 55& -0.0381& -0.0382& 75& -0.0031& -0.0032\\
16& -0.1846& -0.1845& -0.1846& 36& -0.0982& -0.0982& -0.0982& 56& -0.0356& -0.0356& 76& -0.0024& -0.0024\\
17& -0.1797& -0.1797& -0.1797& 37& -0.0945& -0.0945& -0.0945& 57& -0.0332& -0.0332& 77& -0.0017& -0.0018\\
18& -0.1749& -0.1749& -0.1749& 38& -0.0909& -0.0909& -0.0909& 58& -0.0308& -0.0308& 78& -0.0012& -0.0013\\
19& -0.1701& -0.1701& -0.1701& 39& -0.0873& -0.0873& -0.0873& 59& -0.0285& -0.0285& 79& -0.0008& -0.0009\\
  &&&&&&&&&&									  & 80& -0.0005& -0.0006\\
  &&&&&&&&&&									  & 81& -0.0002& -0.0004\\
  &&&&&&&&&&									  & 82& -0.0001& -0.0002\\
  &&&&&&&&&&									  & 83&        & -0.0001\\
  &&&&&&&&&&									  & 84&        & -0.0001\\
%  &&&&&&&&&&									  & 85&        & -0.0000\\
%  &&&&&&&&&&									  & 86&        & -0.0000\\
%  &&&&&&&&&&									  & 87&        & -0.0000\\
\hline\hline
\end{tabular}}
\end{center}
\label{TvSClF}
\end{table}

\section{Discussion}

The complete rovibrational spectra is depicted in the histogram in
Fig.~\ref{rvsCS} with energies larger than $\sim 10^{-4}$\,{\it hartree}. In total,
considering $E^{C_6}$ \re{fitP49}/$E^{C_5}$\re{fitP59} we arrive at 14562/14624
B-O rovibrational states, respectively, which is in reasonably good agreement
with 14908 rovibrational states found in \cite{PCSFBMM:2018}. 
For both cases \re{fitP49} and \re{fitP59} the same maximal angular momentum 
$L_{max} = 289$ is predicted (if weakly-bound states with energies 
$\lesssim 10^{-4}$~{\it hartree} are taken into account), 
contrary to $L_{max} = 258$ \cite{Paulose:2015} and $L_{max} = 260$ \cite{HW:2020}.  
Following the histogram of Fig.2 it implies that $\sim 250$ rovibrational states with
$L=260 - 289$ are missing in analysis of the rovibrational spectrum performed in 
\cite{Paulose:2015,HW:2020}. Thus, this becomes surprising why in \cite{PCSFBMM:2018} 
the number of rovibrational states is larger than ones predicted with \re{fitP59} 
in the present work by a large number $\sim 370$.} 

In general, for each value of the angular moment $L$,  the number of bound states
obtained when considering $E^{C_5}$~\re{fitP59} is greater than or equal to the
number of states when taking $E^{C_6}$~\re{fitP49}: $N^{E^{C_5}}\geq N^{E^{C_6}}$.
Single exception appears related with the weakly bound state $E_{(45,174)}$, which
is present if $E^{C_6}$ ($C_5=0$) case is considered, see \re{fitP49}, but absent 
in $E^{C_5}$ ($C_5\neq 0$) case, see \re{fitP59}.
This state is highlighted in red in Fig.~\ref{rvsCS}.

\begin{figure}[h!]
\includegraphics[scale=2.5]{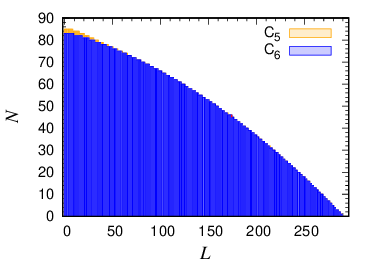}
\caption{B-O rovibrational spectra for the ground state $X^1\Sigma^+$ of carbon
         monosulfide $^{12}$C$^{32}$S with energy $\gtrsim 10^{-4}$\,a.u.  
         Blue states result from solving the nuclear Schr\"odinger equation~\re{NucSc} 
         for the potential~\re{fitP49}
         whose dominant term at large internuclear distances is $\sim C_6/ R^6$
         while those in orange the dominant term is $\propto C_5/ R^5$.
         State $(45,174)$ (in red) is found as bound state when $C_6\neq 0$ ($C_5=0$), 
         see \re{fitP59}, but not found when $C_5\neq 0$ \re{fitP49}.}
\label{rvsCS}
\end{figure}

It must be emphasized that the phenomenological analysis carried out in \cite{CH:2023} takes into account 
the available experimental data for angular momenta $L=0 - 144$. It implies immediately the enormous amount 
of missing rovibrational states in the analysis with higher angular momentum $L > 144$ in \cite{CH:2023}. Other missing (or wrongly predicted) rovibrational states would correspond to larger vibrational quantum numbers $\nu > 39$ at $L > 0$. It indicates that phenomenologically introduced factors in front of kinetic energy and centrifugal potential as well as a reduced mass dependence in the potential (curve) in the radial nuclear Schr\"odinger operator are sufficient to obtain spectroscopic accuracy for the part of the spectrum corresponding the experimentally observed states but completely insufficient to predict correctly so-far-non-observed-experimentally states.  
This reminds the situation with analysis of halides performed in \cite{CH:2015}, where thousands of rovibrational states were missing, see \cite{OT:2023,AdT:2024}. 

\section{Conclusions}
 
Using several turning points on the potential curve found by Coxon-Hajigeorgiou \cite{CH:2023}
in the domain $R \in (2.3, 4.5)$\,a.u. and the 9 turning points in $R \in (5.0,7.5)$\,a.u.,
together with asymptotic behavior given by perturbation series 
in $R$ at small distances and by multipole expansion at large distances $R$ the B-O potential curve 
is built for all $R \in [0, \infty)$ analytically. 
The consistency of the finding turning points is checked by making the comparison of the experimentally-observed transition lines with ones predicted by the spectrum of the radial nuclear 
Schr\"odinger equation with potential given by B-O potential curve (5) or (8).

Knowledge of the B-O potential curve in the whole domain $R = [0, \infty)$ allows us to calculate reliably 
the rovibrational spectrum, both eigenvalues and eigenfunctions, by solving the nuclear Schr\"odinger equation. It gives a chance to calculate reliably the matrix elements and transition amplitudes, in particular, the electric dipole transition probabilities.
The electric dipole transition $(0,0) \rar (0,1)$ will be studied elsewhere.

\section*{Acknowledgements}

The research is partially supported by CONACyT grant A1-S-17364 (in the early stage of the research) and 
DGAPA grants IN113022, IN104125 (Mexico).


\begin{thebibliography}{99}

\bibitem{CH:2023}
         J.A.~Coxon and P.G.~ Hajigeorgiou,\\
%        \textit{The ground electronic state of CS: A global multi-isotopologue
%          direct potential fit analysis},\\
         {\em J. Mol. Spectrosc. \bf 398}, 111861 (2023)\\
          doi.org/10.1016/j.jms.2023.111861

\bibitem{TLO:2019}
         A.V.~Turbiner, J.C.~Lopez Vieyra and H.~Olivares-Pil\'on,\\
%         \textit{Few-electron atomic ions in non-relativistic QED:
%         Ground state energy},\\
         {\it Annals of Physics \bf 409} (2019) 167908 (19 pp)

%\bibitem{Pachucki}

\bibitem{Komasa:2019}
         J.~Komasa et al,\\
%         \textit{Rovibrational energy levels of the hydrogen molecule through nonadiabatic perturbation theory},\\
         {\it Phys.Rev. \bf A 100} (2019) 032519 (10pp)

\bibitem{CH:2015}
         J.A.~Coxon and P.G.~ Hajigeorgiou,\\
%         \textit{Improved direct potential fit analyses for the ground electronic
%         states of the hydrogen halides: HF/DF/TF, HCl/DCl/TCl, HBr/DBr/TBr and HI/DI/TI},\\
         {\em J. Quant. Spectrosc. Radiat. Transf. \bf 151}, 133-154 (2015)

\bibitem{OT:2023}
         H.~Olivares-Pil\'on and A.V.~Turbiner,\\
%         \textit{HCl, DCl and TCl diatomic molecules in their ground state: the
%               B-O rovibrational spectra},\\
         {\it Journal of Physics \bf B56} (2023) 165101 (9pp)
         
\bibitem{AdT:2024}
         Daniel Aguilar-Díaz and Horacio Olivares-Pilón,\\ 
%         \textit{Hydrogen halides HBr/DBr/TBr and HI/DI/TI: 
%         B-O rovibrational spectrum},\\
         {\it Journal of Physics \bf B57} (2024) 215101

\bibitem{WR:1973}
         J.E.~Wollrab and R.L.~Rasmussen,\\
%         \textit{Lifetime of gas-phase carbon monosulfide},\\
         {\em J. Chem. Phys \bf 58}, 4702-4703 (1973)

\bibitem{PSWJ:1971}
         A.A.~Penzias, M.~Solomon, R.W.~Wilson and K.B.~Jefeerts,\\
%         \textit{INTERSTELLAR CARBON MONOSULFIDE},\\
         {\em ApJ \bf 168}, L53-L58 (1971)

\bibitem{MKS:1988}
         E.K.~Moltzen, K.J.~Klabunde and A.~Senning,\\
%         \textit{Carbon Monosulfide: A Review},\\
         {\em Chem. Rev.  \bf 88}, 391-406 (1988)
%\newpage
\bibitem{Paulose:2015}
         G.~Paulose, E.J.~Barton, S.N.~Yurchenko, J.~Tennyson,\\
%         \textit{ExoMol molecular line lists - XII. Line lists for eight isotopologues of CS},\\
         {\em MNRAS \bf 454}, 1931-1939 (2015)

\bibitem{PCSFBMM:2018}
         R.J.~Pattillo, R.~Cieszewski, P.C.~Stancil, R.C.~Forrey, J.F.~Babb, J.F.~McCann
         and B.M.~McLaughlin,\\
%         \textit{Photodissociation of CS from Excited Rovibrational Levels},\\
         {\em ApJ \bf 858}, 10 (2018)

\bibitem{TO:2022}
         A.V.~Turbiner and H.~Olivares-Pil\'on,\\
%       \textit{Towards the analytic theory of Potential Energy Curves for diatomic
%       molecules. Studying He$_2^+$ and LiH diatomics as illustration},\\
         {\it Mol. Phys. 120} (2022) e2064784

\bibitem{OT:2018}
      H.~Olivares-Pil\'on and A.V.~Turbiner,\\
%      {\it H$_2^+$, HeH and H$_2$: Approximating potential curves,
%       calculating rovibrational states},\\
      {\it  Ann. Phys. \bf 393}, 335-357 (2018);
      {\it  ibid  \bf 408}, 51 (2019) (erratum)

\bibitem{OT:2022}
      H.~Olivares-Pil\'on and A.V.~Turbiner,\\
%      \textit{ClF diatomic molecule: rovibrational spectra},\\
      {\it Chemical Physics Letters, \bf 799} (2022) 139642
%       ArXiv: 2202.10666v2, pp.12 (May 2022)
%              DOI: 10.1016/j.cplett.2022.139642

\bibitem{AG-OP:2022}
        L.E.~Angeles-Gantes, H.~Olivares-Pil\'on,\\
%       {\it HF, DF, TF: Approximating potential curves, calculating rovibrational states},\\
%        ArXiv: 2110.01991 (15 pp)
        {\it J Phys \bf B 55} (2022) 65101 (7pp)

\bibitem{NIST:2022}
        A. Kramida, \& Yu. Ralchenko, \& J. Reader \& NIST ASD Team,\\
        \textit{NIST Atomic Spectra Database (ver. 5.10)},\\
        {\em National Institute of Standards and Technology, Gaithersburg, MD.},\\
        Available: https://physics.nist.gov/asd {NIST Atomic Spectra Database} (2022)

\bibitem{Bingel:1958}
         W.A.~Bingel,\\
%         \textit{United atom treatment of the behavior of potential
%         energy curves of diatomic molecules for small $R$},\\
         {\em J. Chem. Phys \bf 30}, 1250-1253 (1958)

\bibitem{MK:1971}
         H.~Margenau and N.R.~Kestner,\\
         {\it Theory of Intermolecular Forces},\\
         {2nd edn, Pergamon Press}, 1971

\bibitem{IK:2006}
         I.G.~Kaplan,\\
         {\it Intermolecular Interactions: Physical Picture, Computational Methods and Model Potentials},\\
         {John Wiley \& Sons}, 2006

\bibitem{HW:2020}
         S.~Hou and Z.~Wei,\\
%         \textit{Line Lists for the X1S+ State of CS},\\
         {\em Astrophys. J. Suppl. S. \bf 246}, 1-12 (2020)

\bibitem{DB:2015}
         D.~Baye,\\
%         {\it The Lagrange-mesh method},\\
         {\it  Phys. Rep \bf 565},   1-107 (2015)

\bibitem{AWT:2003}	
         G.~Audi, A.H.~Wapstra and C.~Thibault,\\
%      \textit{The AME2003 atomic mass evaluation},\\
         {\em Nucl. Phys. A \bf 729}, 337-676 (2003)

\bibitem{MSE:2018}
         V.V.~Meshkov, A.V.~Stolyarov, A.Yu.~Ermilov, E.S.~Medvedev, V.G.~Ushakov and I.E.~Gordon,\\
%         \textit{Semi-empirical ground-state potential of carbon monoxide with physical behavior
%         in the limits of small and large inter-atomic separations},\\
         {\em J. Quant. Spectrosc. Radiat. Transf. \bf 217}, 262–273 (2018)


\end{thebibliography}
\end{document}